\documentclass[12pt]{article}
\usepackage{aaspp4}

\def\wisk#1{\ifmmode{#1}\else{$#1$}\fi}
\def\arcpt{\wisk{''\mkern-7.0mu .\mkern1.4mu}}

\newcommand{\oiii}{[O\,{\sc iii}]}
\newcommand{\oii}{[O\,{\sc ii}]}

\newcommand{\lya}{Ly\,$\alpha$}

\newcommand{\civ}{C\,{\sc iv}}

\newcommand{\gl}{$\lambda$}
\newcommand{\ew}{$W_{\lambda}$}

\slugcomment{To appear in Ap. J. Letters}

\begin{document}

\title{Optical and infrared investigation towards the $z=3.8$
quasar pair PC1643+4631~A~\&~B}
 
\author{Richard Saunders\altaffilmark{1}, 
        Joanne C. Baker\altaffilmark{1}, 
        Malcolm N. Bremer\altaffilmark{2},
        Andrew J. Bunker\altaffilmark{3},
        Garret Cotter\altaffilmark{1},
        Steve Eales\altaffilmark{4},
        Keith Grainge\altaffilmark{1}, 
        Toby Haynes\altaffilmark{1},
        Michael E. Jones\altaffilmark{1}, 
        Mark Lacy\altaffilmark{3}, 
        Guy Pooley\altaffilmark{1}, 
        Steve Rawlings\altaffilmark{3}}
\altaffiltext{1}{Mullard Radio Astronomy Observatory,
  Cavendish Laboratory, Madingley Road, Cambridge CB3 0HE, UK}
\altaffiltext{2} {Leiden Observatory, PO Box 9513, 2300 RA Leiden, The Netherlands}
\altaffiltext{3} {Department of Astrophysics, University of Oxford, Nuclear and
Astrophysics Laboratory, Keble Road, Oxford OX1 3RH, UK}
\altaffiltext{4}{Department of Physics and Astronomy, University of Wales
Cardiff, P.O. Box 913, Cardiff CF2 3YB, UK}
 
\begin{abstract} 
 
In a companion paper (Paper I: Jones et al. 1996) we report the discovery of a cosmic microwave 
background decrement, indicative of a distant cluster with mass
$\sim 10^{15} M_{\sun}$, towards the quasar pair
PC1643+4631~A~\&~B ($z=3.79,3.83$, separation 198$''$).
To search for the cluster responsible, we have obtained $R$-, 
$J$- and $K$-band images of the field and have also carried 
out optical spectroscopy of selected objects in it.
No such cluster is evident in these images. Assuming the cluster causing the decrement is
similar to massive clusters already known, our magnitude limits imply that it
must lie about or beyond $z=1$. This provides independent support for the
X-ray-based distance argument of Paper I. The cluster must 
gravitationally lens objects behind it; for a cluster $z$ around $1-2$, the  Einstein-ring
radius for sources at $z\approx 3.8$ is $\sim 100''$. 
Simple modelling, producing simultaneously 
the S-Z effect and the lensing, shows that the source
positions of quasars A and B lie within $\sim 10''$ of each other 
and may indeed be
coincident. The two quasar spectra are found to be remarkably similar
apart from their one-percent redshift difference. Assuming A and B are
images of
a single quasar, we present a possible explanation of this difference.

\end{abstract}
 

\section{Introduction}
 
In a companion paper (Paper I: Jones et al. 1996), we report
the discovery with the Ryle Telescope (RT) 
of a cosmic-microwave-background (CMB) decrement 
towards the $z\approx 3.8$ quasar pair PC\,1643+4631~A~\&~B. 
We argue there that the decrement is most likely to be due to  
the Sunyaev-Zel'dovich (S-Z) effect caused by a system
with mass $\sim 10^{15} M_{\sun}$ but lying beyond the 
range of X-ray telescopes. 

To further constrain the nature and distance of the system causing 
the CMB decrement, we have 
carried out $R$-, $J$- and $K$-band imaging of the field 
with the optical 4.2-m WHT (William Herschel Telescope, La Palma, Canary
Islands) and the 3.8-m UKIRT (UK Infrared Telescope, Hawaii).
In addition, we have used the WHT to obtain spectrophotometry of quasars
A \& B at higher sensitivity and resolution than previously reported
because, as it proceeded, our investigation increasingly pointed to the
possibility that quasars A \& B are physically related.

\section{Optical and Infrared Observations}

All observations were carried out in the period 1995 June to
August, all in photometric conditions and with a maximum 
airmass of 1.3. The imaging is described first and then the
spectrophotometry. We take
$H_0 = 50$\,km\,s$^{-1}$Mpc$^{-1}$, $\Omega = 1$ and $\Lambda=0$.

$J$- and $K$-band images were obtained with the UKIRT IRCAM3 camera,
on a scale of 0\arcpt 29~pixel$^{-1}$. For $R$-band imaging we used
the Tek~1024 CCD mounted on the Auxilliary Port of the WHT, giving
0\arcpt 22~pixel$^{-1}$ after binning. In each case the useful field
size was $\approx 120\arcsec \times 120\arcsec$. Therefore, in order 
to cover as big a region around the  decrement as practicable, 
we observed in each band the seven fields whose centres are 
given in Table 1. 
Individual $J$ and $K$ frames of each field were dithered 
in a $3 \times 3$ grid with 12\arcsec\ steps  to avoid repeating bad pixels. 
Consecutive grids were offset by 2\arcsec.
For each of fields~0--6, the total
integration time was 27~min in $J$ and 45~min in $K$. 
The infrared seeing was 1--1\arcpt 5 FWHM, but
telescope judder elongated the images to 2\arcsec\ FWHM. In
$R$, a 5-min integration time was used for each of fields 1--6  and 10 min
for field 0; seeing was 1\arcpt 5 FWHM. Images of photometric
standard stars were taken throughout. 

The optical and infrared images were bias-subtracted, flat-fielded
and flux-calibrated using standard procedures in {\sc
iraf}. The limiting magnitudes for point sources
(3-$\sigma$ within 1\arcpt 5 radius)
are as follows: for field~0, $R=24.5$, $J=22.2$, $K=20.6$; for
fields~1--6, $R=24.2$, $J=22.2$, $K=20.6$. 

Optical spectrophotometry was carried out at the WHT 
using the red and blue arms of
the ISIS spectrograph simultaneously. With gratings of 158 line~mm$^{-1}$ 
and Tek CCDs, we obtained continuous coverage from 3500 \AA\ to 8500 \AA\
and a spectral resolution of 12-\AA\ FWHM. A 2\arcsec-wide
slit was used with typical seeing 0\arcpt 9--1\arcpt 2 FWHM. 
Exposures of spectrophotometric 
standards were taken routinely. 
In this way, spectra were 
obtained for quasars A and B, each with an integration time of 25 min, 
and also 
for objects close to the  decrement centre, typically with 10--15~min 
exposures. Bias subtraction, flat-fielding and flux calibration were done
using {\sc iraf}.

\section{Analysis of the Field}

\begin{figure}[t!]
\vspace{1cm}
\plotone{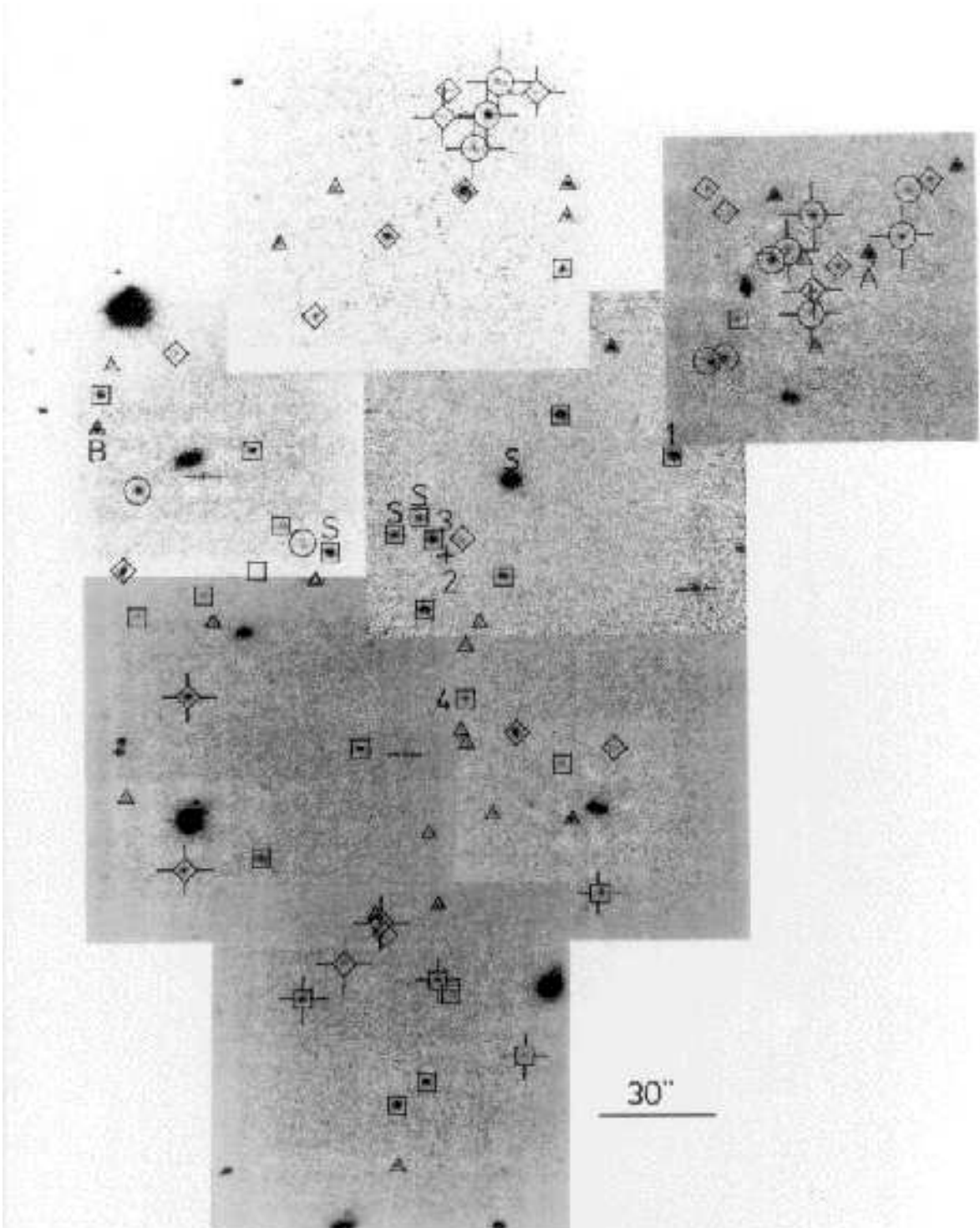}
\newpage
\caption{Mosaic of infrared $K$-band images of fields 0--6 (see text). 
Objects with $K\le 19.5$ have been assigned colour-dependent symbols.
The following $R-K$ ranges are marked:
0.0--2.4 by $\bigtriangleup$, 2.5--3.4 by $\sq$, 3.5--4.4 by {\large$\diamond$}, 
$>4.5$ by {\large$\circ$}. 
A cross is superposed if $J-K>2.0$. Objects marked with 
two horizontal bars have $K>19.5$ but no colour information.
Spectroscopic identifications are also shown for a number of objects:
quasars ``A'' \& ``B''; stars ``S''; galaxies ``1'', ``2'' \& ``3'' 
(see text).  
The centre of the S-Z decrement lies near object ``4''; note that the
positional accuracy of the decrement centre is $\approx20''\times 30''$
(see Paper I). The different apparent background levels between the fields
reflect small variations in sky noise between the observations; 
we have used a compressed greyscale to emphasise low surface brightness
objects. The E-W elongation of the images is due to UKIRT telescope judder.
}
\label{mosaic}
\end{figure}

To search for a candidate cluster, we classified objects in the field by
colour.
We measured magnitudes (integrated over a 3\arcsec-radius aperture)
for all objects detected in $K$ in the seven fields, except for 
stars and bright galaxies which were obviously nearby.
Fig. \ref{mosaic} shows the composite $K$ image. All objects with $K
\leq 19.5$ (the 3-$\sigma$ photometry limit is $K=19.9$) 
are marked with a colour-dependent symbol, except for very bright 
objects and objects at the extreme periphery of the composite.
The form of the symbol is
governed by the redness ($R - K$) of the object. A cross is added if
$(J - K) > 2$, perhaps indicative of a high-redshift elliptical
(e.g. Bruzual \& Charlot 1993). 

The immediate result is that in these deep images no cluster capable of causing the
 decrement is evident. We do see red galaxies with $(J - K) > 2$
both around quasar~A and in the southern part of the composite,
including the red galaxies found by Hu \& Ridgway (1994). 
 If these galaxies with $K< 19$ lie at 
$z\ga 2$, as their colours suggest, they would be as luminous as 
the brightest galaxies in
low-redshift clusters (Arag\'on-Salamanca et al. 1993), i.e. have
luminosities greater than a few $L_{\ast}$. A band of objects can also be
seen across the field, mostly with $2.5 < (R - K) <3.5$. 
We obtained spectra for some of these: the four objects marked ``S'' on 
Fig. \ref{mosaic} turned out to be stars; object 1 ($K = 17.0 $) is an
emission-line galaxy with strong 
\oii\,\gl3727 and H-$\beta$ at $z = 0.659 \pm 0.002$; objects 2 and 3 are
companion galaxies with \oii\ emission at $z = 0.670 \pm 0.002$  
(object 2  is not detected in $J$ or $K$ but has $R =22.2 $; object 3
has $K = 17.4$).

We reiterate that these observations do {\it not\/} reveal 
evidence of a cluster near the decrement position, implying the cluster is
distant. 
>From our knowledge of the cluster luminosity function at $0.5 < z < 1$
(e.g. Arag\'on-Salamanca et al. 1993),
a cluster at, for example, $z =0.7$ should have a brightest member of
$K \approx 16.2$ and several tens of members present 
brighter than $K = 19$. This is clearly not observed ---
recall our 3-$\sigma$ photometry limit over a
3\arcsec-radius aperture is $K=19.9$.  
Following this argument, our images
imply that any ``normal'', S-Z-producing cluster must lie at
about $z = 1$ or beyond. This is consistent  with the X-ray result of Paper I.
The existence of a significant population of such clusters is not
favoured by current ``bottom-up'' models of structure formation
(e.g. cold dark matter, see e.g. Peebles 1993).

Of course, one can speculate that the cluster is, 
compared with known clusters, underluminous in X-rays because the
electron density, $n$, is low, and underluminous in the
optical/infrared because because its galaxies have burned out (see Silk 1986)
or little star formation has occured in it.  However, to produce an S-Z signal, the
line-of-sight pressure integral $\int n T dl $ must be maintained, so a system with mass $\sim 10^{15} M_{\sun}$
is {\em still} required.

\section{The System as a Gravitational Lens}

Given that a massive system must lie in the direction of the
decrement, gravitational lensing will occur. Moreover, if the system
lies at $z < 3.8$ it will affect the apparent (``object'') positions
of quasars A \& B.  To investigate the unlensed (``source'') positions
of A \& B, we have carried out simple modelling subject to the additional
constraint that the lens gas causes an S-Z effect of the observed
magnitude (see Paper I).  We employed the lensing formulation of
Blandford \& Narayan (1992) and a standard
spherically-symmetric ``$\beta$-model'' to describe the S-Z-producing
gas. In this model the density is given by $n = n_0 \{ 1 +
(\theta/\theta_C)^2 \}^{-1.5 \beta}$, where $\theta_C$ is the
core-radius (see e.g.  Cavaliere \& Fusco-Femiano
1976). We have assumed $\beta = 0.67$ and a temperature
of $5 \times 10^7$~K, both typical of known X-ray clusters. As a first
approximation, we placed the lensing mass at $z = 1$, adopted a ratio
of total mass to gas mass of 10:1, and made the lens and S-Z centres
coincident.

By adjusting the values of $\theta_C$ and $n_0$, we determined the
source positions as a function of cluster gas
distribution. Intriguingly, a distribution with $\theta_C = 35''$
(300\,kpc) and $n_0 = 7 \times 10^3$ m$^{-3}$ both produces the S-Z
decrement observed with the RT and makes the source positions of
quasars A \& B {\it almost coincident} (within 10\arcsec). The
Einstein-ring radius is $100''$ and the total mass, including dark
matter, within a 1-Mpc radius of the centre is $1.2 \times 10^{15}
M_{\sun}$.

 It is remarkable that these values of core radius and mass are
 completely typical of known luminous clusters. The values are not
 particularly sensitive to $\beta$ or $T$, nor to the redshift of the
 lens (the mass required changes by less than fifty percent for $0.6 <
 z < 3$).  In other words, $\int n T dl$ is fixed by the S-Z
 decrement, and if one takes gas parameters similar to those of known
 clusters, the gravitational-lensing effects are such that the source
 positions of A \& B must be close together.  Of course, minor
 adjustments to the assumed lens potential could make the quasar
 source positions coincide exactly. Then quasars A \& B would be
 double images of the same object. This would be by far the
 largest-separation multiple-image known. At least one other image
 would be expected but it could easily be fainter than our current
 magnitude limits.

\begin{figure}[t!]
\plotone{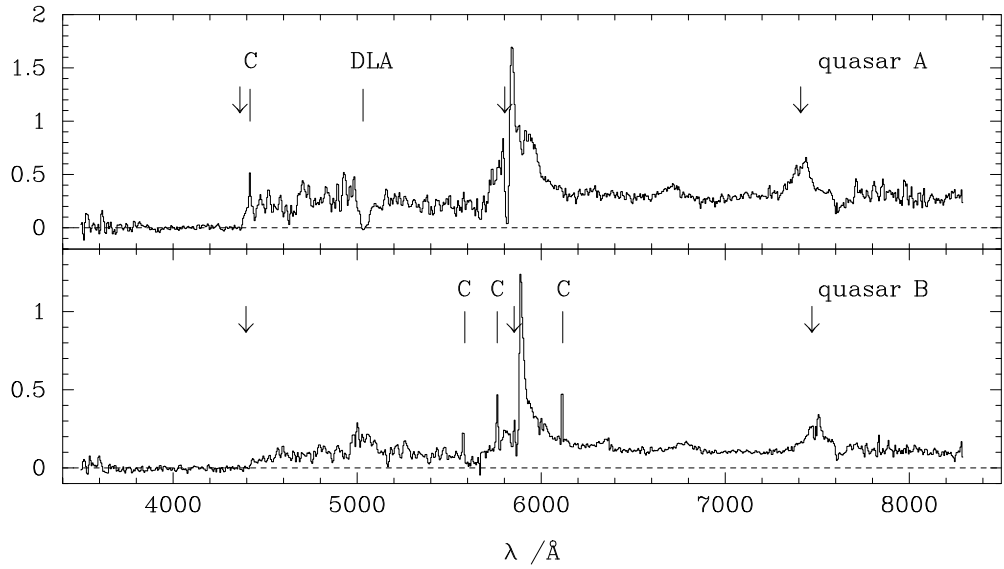}
\caption{WHT spectra for quasars~A and B. The ordinate scales are
$f_{\lambda}$/$10^{-19}$\,W\,m$^{-2}$\AA$^{-1}$. Cosmic rays are marked
``C''; night sky  affects the spectra at 7600\AA\ and to the red.
Associated absorption features are marked in both spectra with vertical 
arrows, including Lyman-limit, Ly\,$\alpha$ and C\,{\sc iv} absorption. 
The spectrum of quasar A also shows damped Lyman-$\alpha$ (DLA) 
absorption at 5032\AA. 
}
\label{qsospectra}
\end{figure}

The possibility that quasars A \& B are the same object compelled us
to compare their spectra as closely as possible.  The WHT spectra we
obtained for quasars A \& B (see Fig. \ref{qsospectra}) have higher
spectral resolution and sensitivity and extend further into the blue
than those published by Schneider et al. (1991). We confirm their
broad-line redshifts and damped Ly-$\alpha$ absorption at
$z=3.14$ in the spectrum of quasar A. 
The continuum shapes, line strengths and widths are similar
in both quasars and two additional features in common are particularly
striking. First, narrow absorption is clearly present in both objects,
in both \lya\,\gl1216 and \civ\,\gl\gl1549,1551, shifted in both
spectra $\sim 6000$~km\,s$^{-1}$ to the blue of the line peaks.
Second, abrupt Lyman-limit absorption is seen in the light from both
quasars at the same redshifts as the narrow absorption features.
Associated Lyman-limit systems are seen in about 20\% of quasars with $z>2.5$
selected for absorption-line studies 
(Lanzetta 1991; Storrie-Lombardi et al. 1994), although the probability of 
detecting two systems with similar relative velocities is much smaller. 
For intervening systems, a comoving number density of 0.06 Lyman-limit
absorbers per line of sight is expected within a redshift interval of 
0.02 at $z = 3.8$ (Storrie-Lombardi et al. 1994). We note that
associated Lyman-limit systems will follow a different redshift 
distribution if they are affected directly by the quasar environment,
which is likely.

However, despite the qualitative similarity of spectral 
features and gaseous environments, the redshift difference 
between quasars A \& B remains significant. We find the redshifts of the
broad emission lines in quasars A \& B to differ by 
$(3.0 \pm 0.5) \times 10^3$~km\,s$^{-1}$, with a similar difference for
the Lyman-continuum and narrow C{\sc iv} absorption, and find a
difference in narrow \lya\ absorption of
$(3.0 \pm 0.3) \times 10^3$~km\,s$^{-1}$.

For quasars A \& B to be two gravitationally-lensed images of one
source, their one-percent redshift difference has to be explained by 
intrinsic spectral changes occuring over the likely
delay between the two lightpaths,  $\sim 10^3$ years.
We note that velocity shifts, comparable with those observed 
between quasars A \& B, have been observed between high- and low- ionisation 
emission lines in numerous quasar spectra and interpreted as 
evidence for radial velocities 
within the broad-line region (BLR) (see Gaskell 1982; Wilkes 1986). 
Such a radial, or bulk, velocity that changes over $\sim 10^3$
years would provide an explanation for the velocity differences of quasars
A \& B; it is reassuring that the bulk-velocity changes of 3000~km\,s$^{-1}$
are less than the random spread (\lya\ and \civ\ have FWZI of
$(21\pm 3) \times 10^3$ and $(12\pm 2) \times 10^3$~km\,s$^{-1}$,
respectively). 

A model involving a disk (perhaps warped and twisting due to an
interaction) and outflow together with the effects of shielding can
explain the velocity differences {\it provided\/} the shifts of the
narrow absorption features are tied to the shifts of the broad emission
lines. This amounts to a requirement that the absorbing gas lies in or
immediately adjacent to the BLR, rather than at kpc-scale distances in
the narrow emission-line region (NLR). We next argue that this is plausible.

The large optical depth ($\tau > 4$) in the
Lyman-limit absorption trough suggests that both quasars are viewed
through a neutral hydrogen column of at least $10^{21}$~m$^{-2}$ 
(e.g. Sargent 1988). Furthermore, the measured equivalent 
width\footnote{The relative strength of these
absorption lines is affected by additional uncertainties due to the 
complex absorption in the blue wing of the \lya\ emission.}   
of the strongest absorption feature in the \lya\ emission line,
\ew$=20\pm 3$\AA\ in quasar~A ($13\pm 2$\AA\ in quasar~B), corresponds
to a range of column densities $10^{24}<N_{\rm H\,{\sc
i}}<10^{27}$~m$^{-2}$ using a curve-of-growth analysis (e.g.~Spitzer 1978).
Taking a representative BLR density of 
$n_{\rm H\,{\sc ii}}\sim 10^{15}$~m$^{-3}$ (e.g.~Osterbrock 1993),
assuming the neutral hydrogen has the same temperature as the BLR and
is in pressure balance with it and has a filling factor of $10^{-3}$, 
these column densities correspond to line-of-sight 
distances of $10^{11}$--$10^{15}$~m, i.e. much less than a parsec.
Taking a spherical geometry, the corresponding upper limit for the 
H\,{\sc i} mass is $\sim 0.01 M_{\sun}$. These parameters suggest the amount
of material needed to absorb \lya\ in the quasar is easily
containable within the BLR.

A critical test of the notion that the spectra of A \& B are of the same
quasar seen at different times would be to see if narrow lines such as
\oiii\,\gl5007 (in the infrared) lie at the same redshift in both spectra; 
such lines from the large NLR will not vary over $\sim 10^3$ years.

\section{Conclusions}

(1) $R$-, $J$- and $K$- imaging of the field around PC\,1643$+$4631~A~\&~B
has not revealed a cluster in the direction of the CMB decrement. Therefore,
if the $10^{15}$-$ M_{\sun}$ cluster producing the decrement is similar to known massive clusters, it must lie at about $z=1$ or beyond. This is consistent with
the X-ray-based argument of Paper I.

(2) If the system causing the decrement lies
significantly nearer than $z=1$, it must contain far fewer
luminous galaxies than the nearer clusters known to cause S-Z effects. 
It must also be larger, hotter
and more rarefied than known clusters in order to give a lower $\int
n^2 dl$ (and hence lower X-ray luminosity) and yet give the observed
S-Z decrement.

(3) If the massive 1643$+$4631 system is a member of a significant population,
then neither the higher-$z$ picture in (1) nor the lower-$z$ picture in
(2) appears consistent with current cold-dark-matter theories of cluster
formation.

(4) Whether (1) or (2) applies, the
$10^{15} M_{\sun}$ required to produce the decrement must
gravitationally lens background objects. Assuming the mass 
lies in front of the quasars, simple modelling, producing simultaneously 
the S-Z effect and lensing, shows that
the source positions lie within 10\arcsec\ of each other.  

(5) A slight adjustment to the assumed lens potential would make the 
quasar source positions coincide. 
If A \& B are two images of one quasar, then the one-percent
difference in broad-emission and narrow-absorption velocities, in
otherwise very similar spectra, has to be explained. We have argued that
this is possible given (a) the expected $\sim 10^3$ year difference in
light travel time along the two paths, and (b) a system with a changing
systematic velocity component in which narrow-line absorption takes place
in the same region as the broad-line emission; we have shown that the
mass and size of the required absorption region are small enough that
this is plausible.

\subsection*{Acknowledgments}

We are grateful to the staff of the WHT and UKIRT; these telescopes are
operated by the PPARC. We thank the anonymous referee for helpful
comments.

\begin{table}
\caption[Observing Log]{Optical/Infrared Field Centres}
\label{obs}
\footnotesize
\begin{tabular}{ccl} \hline 
 Field & RA~~~(B1950)~~Dec & Comments\\ 

 0 & 16 43 42.7 +46 30 49 & Central field\\
 1 & 16 43 41.6 +46 29 39 & \\
 2 & 16 43 46.0 +46 28 24 & \\
 3 & 16 43 48.7 +46 29 39 & \\
 4 & 16 43 35.3 +46 31 23 & includes quasar~A\\
 5 & 16 43 50.4 +46 30 55 & includes quasar~B\\
 6 & 16 43 45.0 +46 32 00 & \\

\hline
\end{tabular}
\end{table}



\end{document}